# Generating and Managing Strong Passwords using Hotel Mnemonic

**Sumith Yesudasan\***

Department of Engineering Technology, Sam Houston State University, Huntsville, TX, USA

\*Corresponding Author: sumith.yesudasan@shsu.edu

**Abstract**

Weak passwords and availability of supercomputers to password crackers make the financial institutions and businesses at stake. This calls for use of strong passwords and multi factor authentication for secure transactions. Remembering a long and complex password by humans is a daunting task and mnemonic has helped to mitigate this situation to an extent. This paper discusses creating and using long random password and storing them securely using a hybrid strategy of hash-encryption method. The hash function uses a mnemonic password based on the hotel names and other characteristics like room number, floor number and breakfast meal preferences to generate the encryption key. The random strong password can be then encrypted using the key and stored safely. A computer program named Hector is developed which demonstrates these steps and can be used to generate and store the passwords.

**Introduction**

A password, sometimes called a passcode, is a secret string of characters used to confirm someone's identity. Traditionally, passwords were expected to be memorized, but with the increasing complexity of the passwords and need for multiple passwords for various services poses a challenge to memorize them. When the user creates and uses an easier password, then it becomes easier for an attacker to guess. However, if the user chooses a set of passwords that are difficult to remember, then it will be inconvenient due to frequent password resets, tempting to use the same password across many accounts, and may pose security threat since these passwords has to be written down somewhere or to be stored. Cyber security experts advice that a stronger password must "have a mix of uppercase and lowercase letters and digits" or "change it quarterly". Also, they suggest that the longer passwords provide more security than shorter passwords with a wide variety of characters since entropy increases with increasing number of password characters.

In a study, the authors found that selecting passwords based on thinking of a phrase and taking the first letter of each word are just as memorable as naively selected passwords, and is just as hard to crack as randomly generated passwords [1]. Combining two or more unrelated words and altering some of the letters to special characters or numbers is another good method [2]. Having a personally designed algorithm for generating obscure passwords is another good method. A mnemonic password is one where a user chooses a memorable phrase and uses a character (often the first letter) to represent each word in the phrase [3]. Asking users to use "both letters and digits" will often lead to easy-to-guess substitutions such as 'E' → '3' and 'a' → '@', substitutions that are well known to attackers. Similarly typing the password one keyboard row higher is a common trick known to attackers. Moreover, the mnemonic phrase-based passwords are getting weaker since the attackers can build the dictionary with possible phrases a person can remember, like a quote from a book, combination of pet's name, relatives' birthday, and age etc. This poses a challenge, and, in this paper, I put forward a strategy to use a hybrid method of mnemonic passwords and randomly generated passwords.

According to ahla.com, every day in USA approximately 2.5 million guests are checked into hotels (considering an occupancy rate of 50%). Most of the hotels share the similar concepts for check-in and infrastructure, and these characteristics can be used to remember passwords. The idea is to remember a favorite hotel name from the list of all hotels (1214 across all states) in the USA listed in the Wikipedia, the floor and room number of your stay, breakfast meal, topping and beverage preference from the list of most popular breakfast foods in USA. This information will be used to lookup a table and converts into a passkey. This secret passkey is then used to encrypt a long and strong random password into a cipher. This encrypted password (cipher) can be stored in a file, or email or simply printed in a paper. The user can use the same hotel preferences to obtain the passkey and decrypt the cipher to obtain the original strong random password.

This paper is divided into three main sections. The first section explains the method to create a passkey using the knowledge of a customer checking into a hotel with a certain meal preference. The second section explains the generation of random passwords up to a length of 64 characters. The third section shows an implementation of storing strong random passwords using a combination of secure hash algorithm (SHA) [4] and Advanced Encryption Standard (AES) [5]. I have created a software which demonstrates the contents of this paper and is named after the Greek hero "Hector".

## 1. Mnemonic passwords using hotel names

Let us consider an example in which one person is checking in to the hotel. Then a combination of hotel names (1214 options, 4 characters), floor number (1000 options, 4 characters), room number (500 options, 3 characters), breakfast meal (61 options, 2 characters), breakfast topping (14 options, 2 characters) and breakfast beverage (11 options, 2 characters) is available to create a 17-character password. Here, the characters represent the integer corresponding to the selections. For example, the first selection is "1717 Broadway", which corresponds to "0001". Each of these characters are converted to a passkey using a list of 93 symbols consisting of alpha-numerals and US keyboard symbols. This alone can create 93^17 possibilities for passkey characters with an entropy, H = 111.165 bits. This is extended further by giving options for the user's spouse, dad, mom, daughter, and son to check-in which together makes 93^102 passkey possibilities with an entropy of H = 666.994 bits. Here, entropy is estimated using the formula $H=\log_2(N^L)$ (Appendix 1 for details). The graphical user interface (GUI) of the Hector software for this module is shown in the Fig. 1. Throughout this paper, we use the term *passkey* to refer the mnemonic phrase which is encryption key and the *password* to the actual random password of the user.

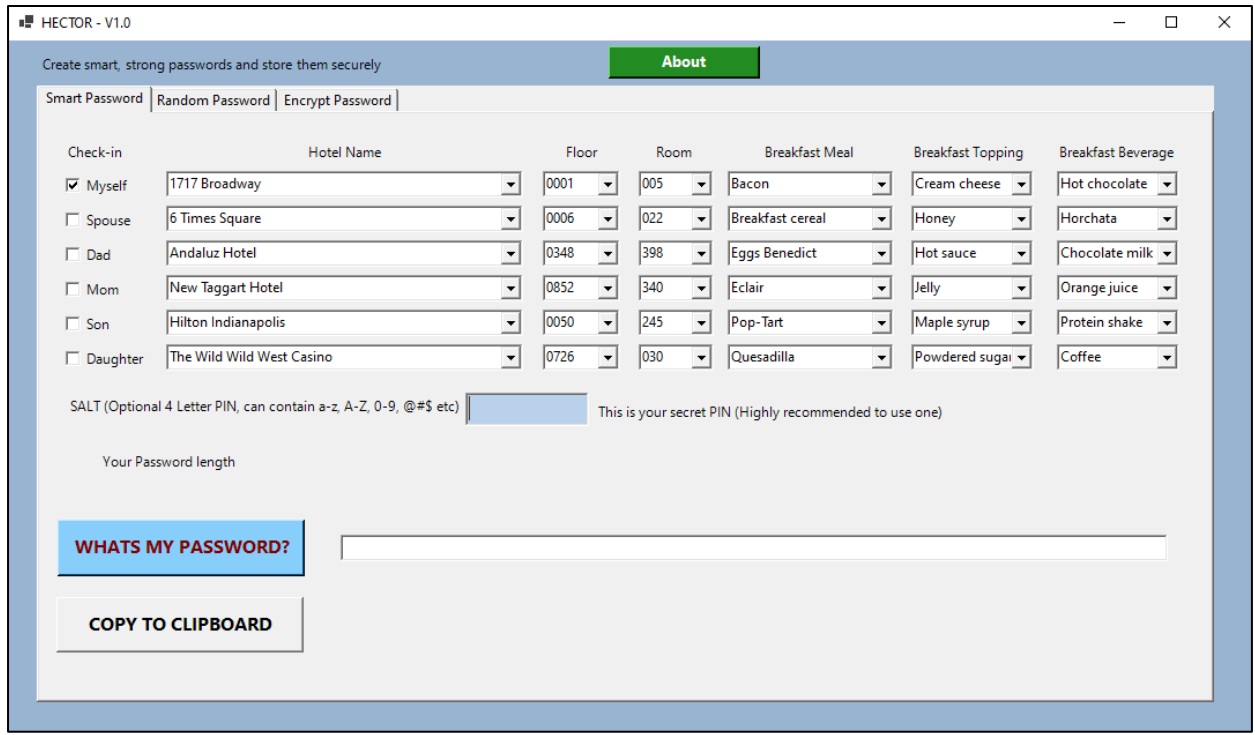

**Figure 1**: GUI of the Hector software's smart password module.

The list of 93 symbols consists of a-z, A-Z, 0-9, and "!#$%&'()*+,-./:;<=>?@[\]^_`{|}~". This symbol list and with the available character combinations, it is not difficult to create a lookup table dictionary of passwords by password cracker. This is a potential security flaw and can be reduced to an extent by adding a "salt". In the first version of this

software, I have used a 4-letter salt, known as PIN to the user which mixes the original selection and makes it difficult for predicting the password using a lookup table. The reason for limiting the salt to 4 letters is to reduce the difficulty to remember the password at the same time preserving some complexity to the password. Having said that, a longer salt length will make the final password more secure from Brute Force and dictionary-based attacks.

**Password estimation steps**

Let us consider for example, the user has selected a hotel (549[th] from the list), floor (128), room (449), breakfast preferences from the list as Bacon (01), Cream cheese (03), Hot chocolate (04). This selection in the Hector software corresponds to an integer sequence as "05490128449010304", which we call it as the initial *passkey*. The *dictionary* is a long phrase with nonrepeating alphabets. The *salt* is a 4-letter word which can be a combination of letters, numbers, and symbols. The steps to convert this information into an alphanumeric passkey is shown below.

<u>Step 1</u>: Convert *salt* to an integer using the formula, $X_0 = \sum_{j=1}^{4} F_{ASCII}(salt[j]) 10^j$

<u>Step 2</u>: Select the i[th] character from the predefined *dictionary* of length $L_2$, convert it into integer and let's call it as $D$. i.e., $D[i] = F_{ASCII}(dictionary[i])$.

<u>Step 3</u>: Define the symbol list used to make passkey as *symbol list* (SL) with a length $L_3$, which consists of a-z, A-Z, 0-9, and the keyboard symbols.

Step 4: Calculate $X_1 = (m \times X_0 + k) \% L_2$. Here, $m$ is an integer multiplier and $k$ is an integer increment. This operation is called linear congruential generator and is derived from the pseudo random number generator. The symbol "%" represents the modulo operator.

Step 5: Select character by character from the user integer sequence and perform the operation as $PASSWORD[i] = SL[(X_1 + UP[i] + D[i]) \% L_3]$

A pseudocode for the above steps is given below.

```
1  void getpassword (string salt, string passkey)
2  {
3    L1 = get_length_passkey (passkey);      //05490128449010304
4    L2 = get_length_dictionary (dictionary);   //nonrepeating alphabets
5    L3 = get_length_symbollist (symbollist);   //a-z, A-Z, 0-9, @#$ etc.
6    // necessary condition of L1 < L3 <= L2
7
8    for (int j = 1; j <= 4; j++)
9      {
10        X0 = X0 + convert_to_ascii (salt[j]) * math.pow (10, j));
11     }
12
13    X1 = (m * X0 + k) % L2;
14
15    for (int i = 1; i <= L1; i++)
16      {
17        UP = strip_user_pass (passkey, i);    //get the ith character
18        D = convert_to_ascii (dictionary[i]);
19        index = (X1 + UP + D) % L3;
20        passkey [i] = symbollist[index];
21     }
22 }
```

This will yield "3_`pV1_G)z56!uD8f" as the passkey with a salt of "45%D", $m = 45$ and $k = 241$. If the user doesn't supply a salt/PIN, then the software will use a default salt of "7391". Here, the default values for $salt$, $m$ and $k$ are chosen arbitrarily and will be changed for the release version of the Hector software. By this way, the user can create a strong passkey by remembering a PIN and the hotel configuration. The unknown parameters to the attackers are the $salt$, $m$, $k$, $L_2$, and $dictionary$.

## 2. Random password generation

At times it is necessary to use a password which is completely random which doesn't have a pattern or constructor that can be traced to. This section explains the basic way to create such random password in Hector software. The GUI of Hector software is shown in the Fig. 2 below and user has the options to select the length of the password, type of the combinations for the character and symbols for it etc.

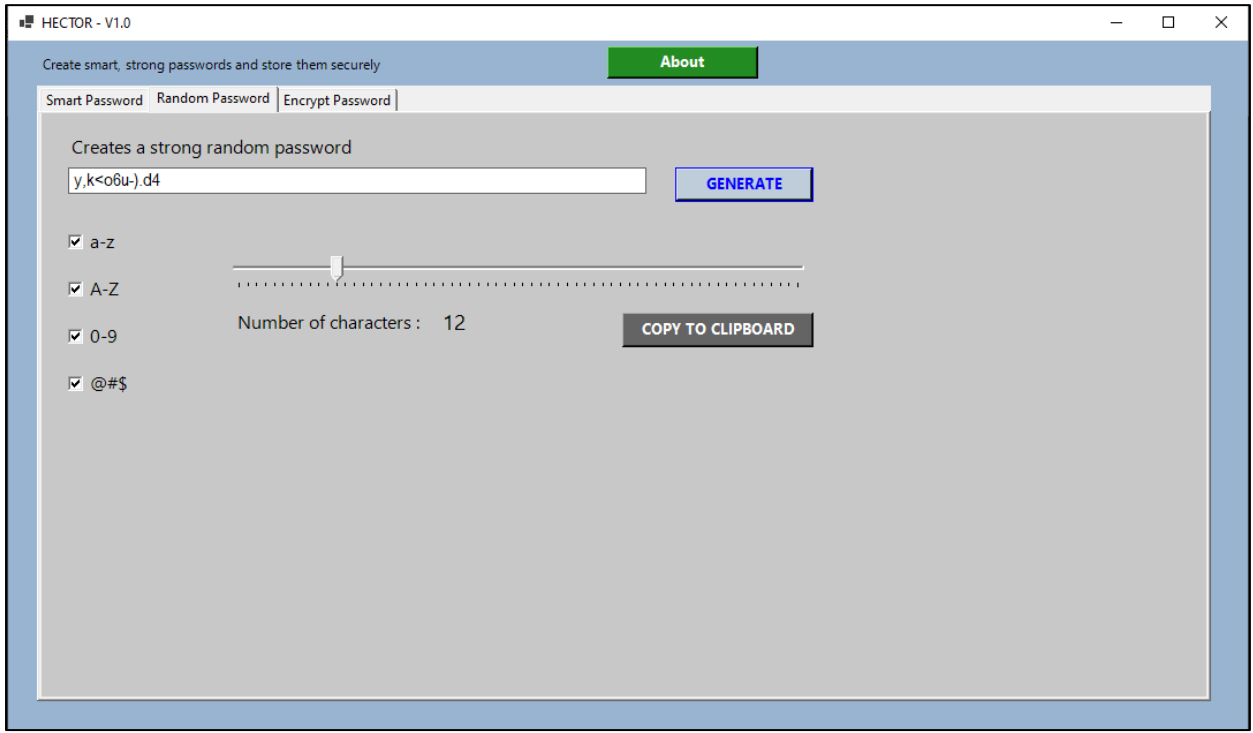

**Figure 2**: GUI of the Hector software's random password module.

The character list used here consist of letters from a to z, A to Z, numbers from 0 to 9, and the symbols "!#$%&'()*+,-./:;<=>?@[\]^_`{|}~". This corresponds to a maximum possible character set of length 93. It is known fact that creating true random numbers in software is a challenge and hence we rely on the pseudo random number generator (PRNG) for this purpose. Hector software uses the VB Net *rnd* function which is a linear congruential generator (LCG). The working steps of this implementation is as explained below.

Step 1: Obtain the number of characters ($L_1$) required for the password. Obtain the set of character list from the options "a-z", "A-Z", "0-9" and "@#$".

Step 2: Construct a string sequence with the character list set chosen. Let $L_2$ be the character length of this string.

Step 3: Create a random integer number $N$ between 0 and $L_2$.

Step 4: For each required characters of the password length, find the character from the position $N$ from the list generated in the step 1.

Step 5: Repeat step 3 and step 4 until the iterator becomes $L_1$.

Some of the sample random passwords generated are "JG3/_s&)0B*=", "lC(>[U-e2e5W0h/BPl@5", "[5dT(k;XTSk>JI~,l{}9" etc.

Pseudocode for the random password creation is given below.

```
1  string randompassword(int L1)
2  {
3   string password = "";
4   string dictionary = createdictionary();
5   int L2 = length(dictionary);
6   for(int i = 1; i <= L1; i++) {
7      N = randombetween(0 and L2);
8      password = password & dictionary[N];
9   }
10    return password;
11 }
12 string createdictionary() {
13    string dictionary = "";
14    if(option == 'a - z')
15     dictionary = dictionary & 'a…..z';
16    if(option == 'A - Z')
17     dictionary = dictionary & 'A…..Z';
18    if(option == '0 - 9')
19     dictionary = dictionary & '0…. .9';
20    if(option == '@ #$')
21     dictionary = dictionary & '@…..#';
22    return dictionary;
23 }
```

## 3. Secure storage of passwords

The previous two sections explained about generating passwords based on a mnemonic and in a random manner. In this section, we will use the encryption methods combined with hashing functions to securely store and retrieve passwords. The Fig. 3 shows the GUI of the secure encryption-decryption module of the Hector software. The encryption panel of the Hector consists of two input boxes to collect the user's password and a secret key. Using this secret key, the software will convert the user's password to an encrypted password. That encrypted password is secure according to the current NIST standards and can be store safely in email message or write it in a paper or in a file. Even if the password cracker gets the encrypted password, it is difficult or impossible to reverse calculate the user's password using the current computing facilities in the world. The user's password be retrieved from the stored/saved encrypted password only with the knowledge of encryption key. This process is called decryption and the panel for decryption in the Hector (as shown in the Fig. 3) performs this job. The logical process and other implementation details of this process is explained next.

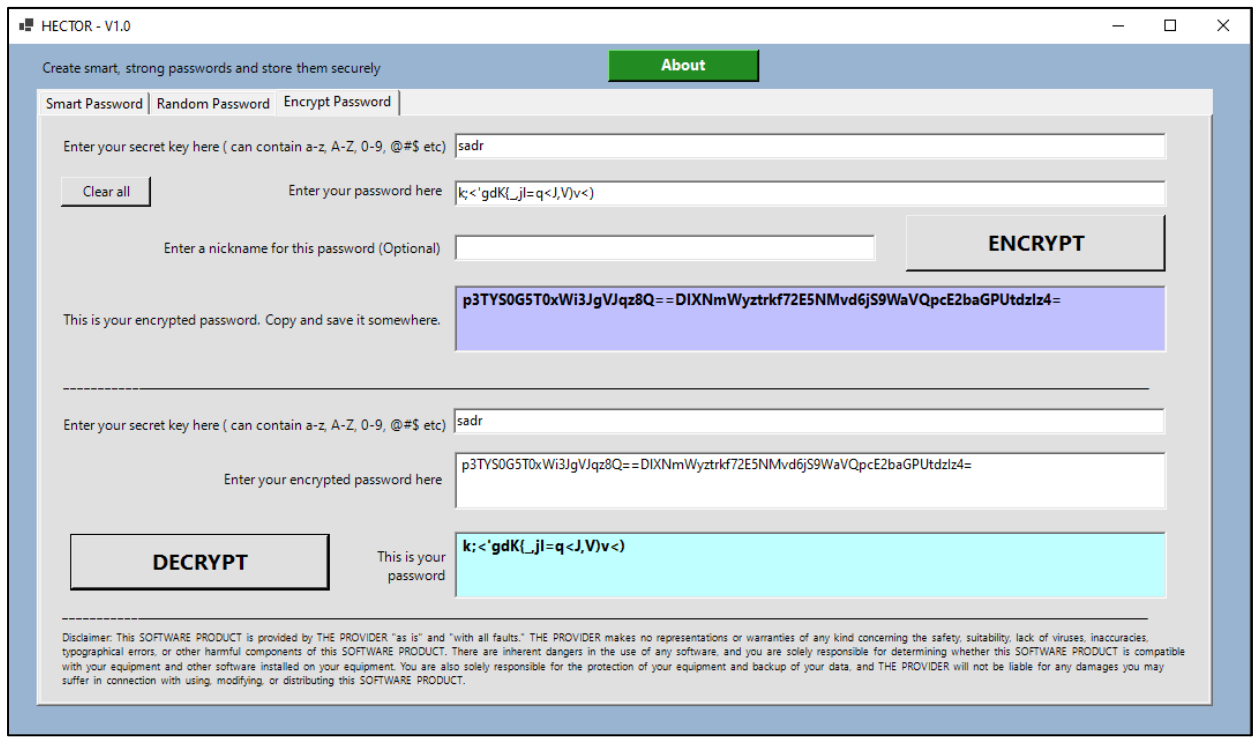

**Figure 3**: GUI of the Hector software's encrypt-decrypt module

## Password storage

The standard encryption algorithms can encrypt the data to a secure form which cannot be reversed without the knowledge of the encryption key. However, to say these algorithms are secure, one must use a random and strong encryption key. This poses a challenge because now we need to protect the encryption key which protects the encrypted data. Otherwise, the user must remember a long random key to protect the password which kills the purpose of a password manager. This is where the combination of encryption algorithm with cryptographic hashing function can help. The password managers can first create an encryption key using the hashing function and convert the complex plaintext password into a cipher using that key. One of such method is explained below, which is used in the Hector software.

## Hash-Encryption using AES and SHA

There are many algorithms available to encrypt the plaintext to a cipher text. From them, I have selected Advanced Encryption Standard (AES) which is also known as Rijndael block cipher due to its ease in implementation with VB Net programming language and compatibility with hash functions. There exist other similar and stronger methods for encryption, but a discussion on that is beyond the topic of interest of this paper. The Advanced Encryption Standard (AES) is published by the National Institute of Standards and Technology (NIST) in 2001. AES is a symmetric block cipher that is intended to replace DES as the approved standard for a wide range of applications. Figure 4 shows the overall process of the AES standard. I have used a key size of 256 bits in this work and the corresponding standard is called AES 256. Using this standard, the plaintext block size is 128 bits, number of rounds is 14, round key size is 128 bits, and the expanded key size is 240 bits. Readers are advised to refer the NIST standard [6] for detailed explanation of implementation and terminology definitions.

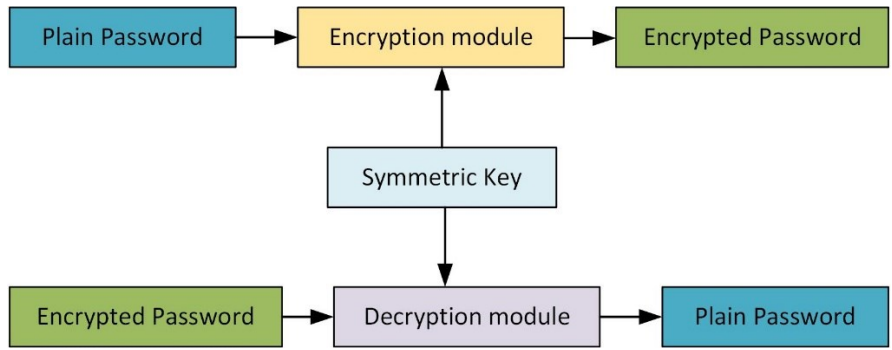

**Figure 4**: Top level AES based encryption and decryption of a password

Due to the secure nature of the AES 256, it is often called as "military grade" encryption algorithm by the crypto community and to date there are no known successful cryptanalysis attacks on AES 256. However, if we use AES 256 alone for encrypting the password and store it, then a bigger problem is storage of the symmetric key which must be complex and long. The user must write the key down or must store it secretly to avoid security issues. This can be solved to an extent by using a cryptographic hash function to generate the key. A cryptographic hash function (CHF) is a mathematical algorithm that maps data of an arbitrary size to a bit array of a fixed size. It is a one-way function which is practically infeasible to invert or reverse the computation. Ideally, the only way to find the data that produces a given hash is to attempt a brute-force search of possible inputs to see if they produce a match or use a rainbow table of matched hashes. I have used the secure hash algorithm (SHA 256) for this purpose.

The steps involved in converting the plaintext password to an encrypted password is shown in the Fig. 5a. The user should supply a secret key to the SHA256 module which generates a hash value. This hash value can be used as an encryption key for converting the plaintext password to a cipher text using the AES256 module. This cipher text can be stored in the files or in emails or can be written down. The original plaintext password can be retrieved by following the steps shown in Fig. 5b. The user creates the encryption key from the secret phrase using the SHA256 module. This key will decrypt the encrypted password using the AES256 decryption module.

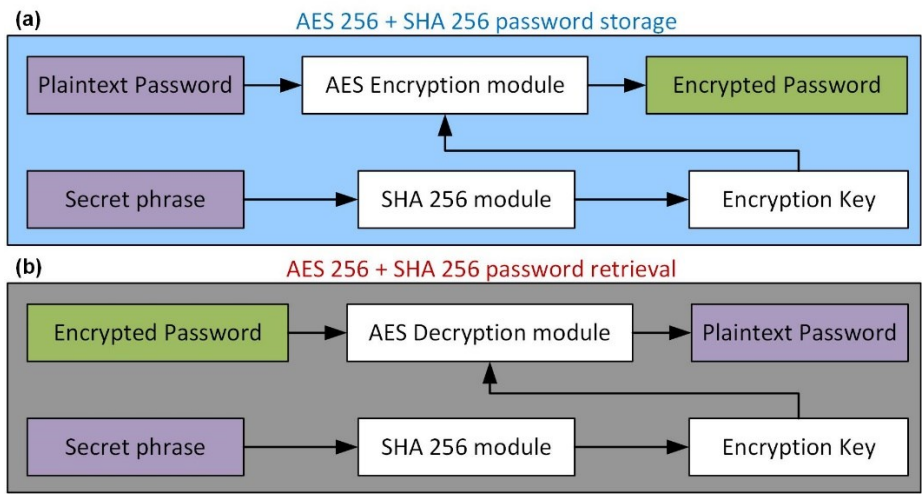

**Figure 5**: a) Steps involved in converting a plaintext password to a secure encrypted password using SHA 256 and AES 256 standards. b) Retrieving steps of the encrypted password using the SHA-AES hybrid method.

Even though the hybrid method of using AES and SHA to store the password looks secure, the strength of this security depends on the length and complexity of secret phrase. If the user is relying on most widely used passwords as the secret phrase, then the attacker can simply brute force it or use a rainbow table to crack it. To avoid this, user must remember a long and strong secret phrase, which is challenging. In these circumstances, the Hector software can combine its three modules to create a strong and secure password storing strategy. Figure 6 explains the top-level working process of this method to store and retrieve complex passwords securely. First, use the random password module of the Hector to create a long and random password to use in banking, emails, logins etc. Next, using the smart password module of the Hector, create the secret phrase or the key. Be sure to use the salt, which adds more security to the process. Finally, use the hotel (smart) password and the random password to encrypt it into a cipher. Store this somewhere handy.

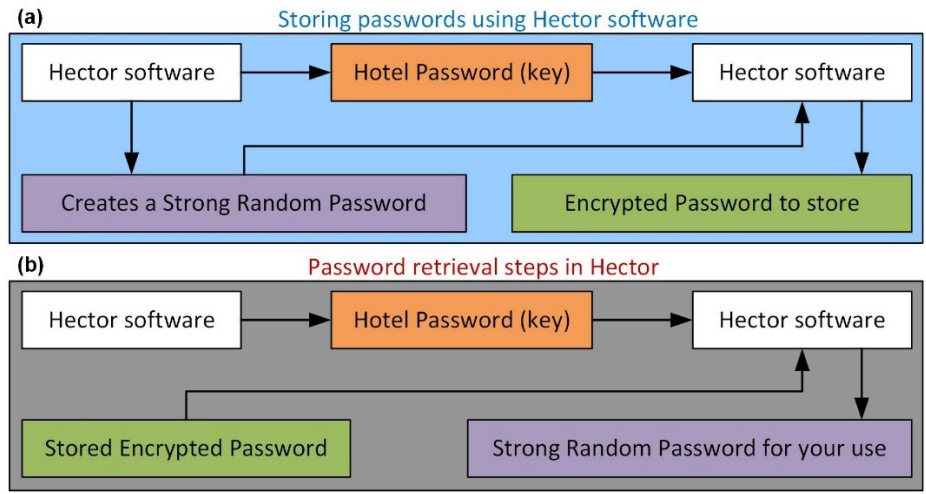

**Figure 6**: a) Suggested steps involved in creating a user password and storing it using Hector software. B) Steps involved in retrieving the stored password using the Hector software.

The stored encrypted password can be decrypted and retrieve the original password using the process shown in Fig. 6b. First create the hotel password using the Hector. Then, supply it as the key and use the Hector's encrypt module to retrieve the random password to use.

## Conclusion

This paper discusses the potential to use hotel names and user preferences while checking into a hotel to remember a mnemonic passkey. This passkey is used to generate a hash function, which is used to encrypt a strong random password. The encrypted password can be stored in files, emails or can be printed. This cryptic form of storage will be difficult for a password cracker to guess the hash function (encryption key) and to decrypt the password. This work also supplements a software named *Hector*, which is implemented with these steps and can be used to store passwords safely.

## Software availability

The Hector software is currently available at free of cost to the public. The executable files are hosted in GitHub and on the author's personal webpage.

Link1: https://github.com/ydsumith/Hector

Link2: https://www.sumith.info/software_1

## Appendix 1: Entropy estimation for smart password.

This section explains the way in which entropy is estimated for the smart password module.
Case 1: Only one person is selected in Hector
    Number of possible symbols, $N = 93$
    password length, $L = 17$
    Entropy, $H = \log_2(N^L)$

        H = 111.165 bits

Case 2: Six people checking into the hotel

        Number of possible symbols, N = 93

        password length, L = 17x6

        Entropy, $H = \log_2(N^L)$

        H = 666.994 bits

## References

[1] J. Yan, A. Blackwell, R. Anderson, and A. Grant, "Password memorability and security: Empirical results," *IEEE Secur. Priv.*, vol. 2, no. 5, pp. 25–31, 2004.

[2] M. E. Whitman and H. J. Mattord, *Principles of information security*. Cengage learning, 2011.

[3] C. Kuo, S. Romanosky, and L. F. Cranor, "Human selection of mnemonic phrase-based passwords," in *Proceedings of the second symposium on Usable privacy and security*, 2006, pp. 67–78.

[4] N. FIPS, "180-2: Secure Hash Standard, August 2002," *URL Httpcsrc Nist Govpublicationsfipsfips180-2fips180-2withchangenotice Pdf*.

[5] J. Daemen and V. Rijmen, "Reijndael: The Advanced Encryption Standard.," *Dr Dobbs J. Softw. Tools Prof. Program.*, vol. 26, no. 3, pp. 137–139, 2001.

[6] V. Rijmen and J. Daemen, "Advanced encryption standard," *Proc. Fed. Inf. Process. Stand. Publ. Natl. Inst. Stand. Technol.*, pp. 19–22, 2001.